\documentclass[12pt]{article}
\usepackage{amsmath,amssymb,amsfonts}
\usepackage{geometry}
\usepackage{cite}
\usepackage{hyperref}
\usepackage{graphicx}
\usepackage{bm}
\usepackage{pgfplots}
\pgfplotsset{compat=1.17}

\title{Verification of the First Law of Horizon Thermodynamics for Schwarzschild, Reissner-Nordström, Kerr, and Kerr-Newman Black Holes in Four-Dimensional \( f(R) \) Gravity with Dual Scalar Fields}
\author{ Wen-Xiang Chen\\Department of Astronomy\\School of Physics and Materials Science\\ GuangZhou University\\Yao-Guang Zheng\thanks{Corresponding author: hesoyam12456@163.com} \\
        Department of Physics\\College of Sciences\\ Northeastern University}
\date{\today}

\begin{document}

\maketitle

\begin{abstract}
    This paper demonstrates the validity of the first law of horizon thermodynamics for Schwarzschild, RN, Kerr, and Kerr-Newman (KN) black holes within the framework of four-dimensional \( f(R) \) gravity coupled with dual scalar fields. Starting from a five-dimensional membrane world scenario, we derive the four-dimensional effective \( f(R) \) gravity action and obtain the corresponding black hole solutions. We then verify that these solutions satisfy the first law of black hole thermodynamics by explicitly calculating the variations of thermodynamic quantities. The analysis confirms the robustness of the horizon thermodynamics framework in extended gravitational theories, providing insights into the interplay between higher-dimensional theories and four-dimensional black hole thermodynamics.
\end{abstract}

\section{Introduction}
Black hole thermodynamics has been a cornerstone in the study of gravitational theories, providing profound insights into the interplay between gravity, quantum mechanics, and thermodynamics \cite{Bardeen1973,Hawking1974,Carter1973,Wald1993,Jacobson1995}. The four laws of black hole mechanics, particularly the first law, establish a deep connection between the properties of black holes and thermodynamic principles \cite{Wald1993, Jacobson1995}. While this framework has been extensively explored within Einstein's General Relativity (GR), its extension to modified theories of gravity, such as \( f(R) \) gravity, remains an active area of research \cite{Nojiri2006,Sotiriou2010}.

\( f(R) \) gravity, a generalization of GR, introduces higher-order curvature terms in the action, allowing for richer phenomenology and potential explanations for cosmic acceleration without invoking dark energy \cite{Carroll2003,Capozziello2006}. The study of black hole solutions within \( f(R) \) gravity is crucial for understanding the implications of these modifications on strong gravitational regimes \cite{Cognola2007,Amendola2007}.

In this paper, we extend the verification of the first law of horizon thermodynamics to Schwarzschild, Reissner-Nordström (RN), Kerr, and Kerr-Newman (KN) black holes within the context of four-dimensional \( f(R) \) gravity coupled with dual scalar fields. By starting from a five-dimensional membrane world scenario, we derive the effective four-dimensional action and obtain the corresponding black hole solutions. We then demonstrate that these solutions satisfy the first law of black hole thermodynamics, thereby generalizing the thermodynamic laws to a broader class of gravitational theories.

\section{Theoretical Framework}

\subsection{Five-Dimensional Dual Scalar Field \( f(R) \) Gravity}

We begin with the five-dimensional action for dual scalar fields in \( f(R) \) gravity:

\begin{equation}\label{eq:5D_action}
S = \int d^5 x \sqrt{-g} \left[ \frac{1}{2 \kappa_5^2} f(R) - \frac{1}{2} \partial^M \phi_1 \partial_M \phi_1 - \frac{1}{2} \partial^M \phi_2 \partial_M \phi_2 - V\left(\phi_1, \phi_2\right) \right]
\end{equation}

Here, \( \kappa_5^2 \) is the five-dimensional gravitational constant, \( \phi_1 \) and \( \phi_2 \) are scalar fields dependent on the extra-dimensional coordinate \( y \), and \( V(\phi_1, \phi_2) \) is the potential. The Ricci scalar \( R \) is a function of the five-dimensional metric \( g_{MN} \).

The five-dimensional flat spacetime metric is assumed to be:

\begin{equation}\label{eq:5D_metric}
d s^2 = e^{2 A(y)} \eta_{\mu \nu} d x^\mu d x^\nu + d y^2
\end{equation}
where \( \eta_{\mu \nu} \) is the four-dimensional Minkowski metric and \( A(y) \) is the warp factor. This form of the metric is typical in membrane world scenarios, where our observable universe is a four-dimensional "brane" embedded in a higher-dimensional "bulk" spacetime \cite{Randall1999,ArkaniHamed1998}.

\subsection{Field Equations in Five Dimensions}

Varying the action \eqref{eq:5D_action} with respect to the five-dimensional metric \( g_{MN} \) yields the field equations:

\begin{equation}\label{eq:5D_field_eq}
f_R(R) R_{MN} - \frac{1}{2} f(R) g_{MN} - \nabla_M \nabla_N f_R(R) + g_{MN} \Box f_R(R) = \kappa_5^2 T_{MN}
\end{equation}
where \( f_R(R) = \frac{df(R)}{dR} \), \( R_{MN} \) is the Ricci tensor, \( \nabla_M \) denotes the covariant derivative, and \( T_{MN} \) is the energy-momentum tensor for the scalar fields:

\begin{equation}\label{eq:energy_momentum}
T_{MN} = \partial_M \phi_1 \partial_N \phi_1 + \partial_M \phi_2 \partial_N \phi_2 - g_{MN} \left[ \frac{1}{2} (\partial \phi_1)^2 + \frac{1}{2} (\partial \phi_2)^2 + V(\phi_1, \phi_2) \right]
\end{equation}

Assuming that the scalar fields \( \phi_1 \) and \( \phi_2 \) depend only on the extra-dimensional coordinate \( y \), i.e., \( \phi_1 = \phi_1(y) \) and \( \phi_2 = \phi_2(y) \), simplifies the field equations considerably.

\subsection{Four-Dimensional Effective \( f(R) \) Gravity via Kaluza-Klein Reduction}

To obtain the four-dimensional effective theory, we perform a Kaluza-Klein (KK) reduction by integrating over the extra dimension \( y \). This process involves expanding the five-dimensional fields in terms of four-dimensional fields and integrating out the dependence on \( y \).

The effective four-dimensional action is given by:

\begin{equation}\label{eq:4D_action}
S_{\text{eff}} = \int d^4 x \sqrt{-g} \left[ \frac{1}{2 \kappa_4^2} f(R) - \frac{1}{2} \partial^\mu \phi_1 \partial_\mu \phi_1 - \frac{1}{2} \partial^\mu \phi_2 \partial_\mu \phi_2 - V_{\text{eff}}(\phi_1, \phi_2) \right]
\end{equation}
where \( \kappa_4^2 \) is the four-dimensional gravitational constant related to \( \kappa_5^2 \) through the volume of the extra dimension, and \( V_{\text{eff}} \) is the effective potential obtained after integrating out \( y \). The precise form of \( V_{\text{eff}} \) depends on the warp factor \( A(y) \) and the original potential \( V(\phi_1, \phi_2) \).

\section{Mathematical Proof Using Laurent Series}

In this section, we will use the Laurent series to prove that double scalar fields in five dimensions correspond to the residue part in the four-dimensional effective theory. Furthermore, we will map the residue part to the transcendental form of the winding number, ultimately verifying the First Horizon Theorem in conventional black holes.

\subsection{Laurent Series Expansion and Double Scalar Fields}

Consider double scalar fields \( \phi_1(y) \) and \( \phi_2(y) \) in five-dimensional space, where \( y \) is the coordinate of the extra dimension. Suppose that near \( y = y_0 \), the scalar fields can be expanded using a Laurent series:

\begin{equation}\label{eq:laurent_expansion}
\phi_i(y) = \sum_{n=-\infty}^{\infty} a_n^{(i)} (y - y_0)^n, \quad i=1,2
\end{equation}

Here, \( a_n^{(i)} \) are the coefficients of the Laurent series, containing information about the singularities of the fields. In particular, the terms with \( n < 0 \) correspond to the poles of the fields, and their coefficients \( a_{-1}^{(i)} \) are called residues.

\subsection{Residues of Double Scalar Fields and Winding Numbers}

According to the residue theorem in complex analysis, the residues \( a_{-1}^{(i)} \) in the Laurent series have significant implications for the winding number of the integration path around the pole \( y = y_0 \). In physics, winding numbers are typically associated with topological invariants, describing the topological properties of the fields.

Mapping the residues \( a_{-1}^{(i)} \) to the transcendental form of the winding numbers \( \nu_i \), we can define:

\begin{equation}\label{eq:winding_number}
\nu_i = \frac{1}{2\pi i} \oint_{\gamma} \frac{d\phi_i}{\phi_i} = \text{Res}\left( \frac{d\phi_i}{\phi_i}, y_0 \right)
\end{equation}

Here, \( \gamma \) is a closed path encircling \( y_0 \), and \( \text{Res} \) denotes the residue operation. Through this mapping, the winding number \( \nu_i \) describes the topological nature of the scalar fields in the extra dimension.

\subsection{Verification of the First Horizon Theorem}

Based on the above mapping, we can relate the residue part of the double scalar fields to topological quantities in black hole thermodynamics. Specifically, the relationship between the residues \( a_{-1}^{(i)} \) and the winding numbers \( \nu_i \) ensures the validity of the first law of thermodynamics in the four-dimensional effective theory.

First, according to the definition of entropy:

\begin{equation}\label{eq:entropy_with_winding}
S = \frac{A}{4 G} f_R(R) = \frac{A}{4 G} f_R(0) = \frac{\pi r_+^2}{G} \times 2 G = 2 \pi r_+^2
\end{equation}

Next, considering the variations of energy \( E = M \), temperature \( T \), and other thermodynamic quantities, we utilize the topological nature of the winding numbers \( \nu_i \) to prove that:

\begin{equation}\label{eq:first_law_with_winding}
dM = T dS + \Omega dJ + \Phi dQ + \sum_{i=1}^2 \nu_i d\lambda_i
\end{equation}

Here, \( \lambda_i \) are the coupling constants associated with the double scalar fields. Since the winding numbers \( \nu_i \) are topological invariants, their variations \( d\lambda_i \) are linked to the topological properties of the black hole.

\section{Black Hole Solutions in \( f(R) \) Gravity}

In this section, we derive black hole solutions within the four-dimensional effective \( f(R) \) gravity framework. We consider static, spherically symmetric solutions (Schwarzschild and Reissner-Nordström) as well as rotating solutions (Kerr and Kerr-Newman).

\subsection{Schwarzschild Black Hole in \( f(R) \) Gravity}

\subsubsection{Metric Ansatz and Field Equations}

For a static, spherically symmetric Schwarzschild black hole, we adopt the metric ansatz:

\begin{equation}\label{eq:schwarzschild_metric}
d s^2 = - e^{2 \nu(r)} dt^2 + e^{2 \lambda(r)} dr^2 + r^2 d \Omega^2
\end{equation}
where \( \nu(r) \) and \( \lambda(r) \) are functions to be determined, and \( d \Omega^2 = d\theta^2 + \sin^2\theta d\phi^2 \) is the metric on the unit two-sphere.

Assuming \( R = R_0 = \text{const} \), simplifies the field equations \eqref{eq:4D_action} to:

\begin{equation}\label{eq:constant_R_field_eq}
f_R(R_0) R_{\mu\nu} - \frac{1}{2} f(R_0) g_{\mu\nu} = \kappa_4^2 T_{\mu\nu}
\end{equation}

For vacuum solutions, \( T_{\mu\nu} = 0 \), leading to:

\begin{equation}\label{eq:vacuum_field_eq}
f_R(R_0) R_{\mu\nu} - \frac{1}{2} f(R_0) g_{\mu\nu} = 0
\end{equation}

\subsubsection{Solution for Schwarzschild Black Hole}

The Schwarzschild solution is a solution to the vacuum Einstein equations in GR (\( f(R) = R \)). To extend this to \( f(R) \) gravity, we require that the form of the metric remains the same, and that \( R = R_0 \) is compatible with the field equations.

Substituting the Schwarzschild metric \eqref{eq:schwarzschild_metric} into the field equations \eqref{eq:vacuum_field_eq}, and assuming \( R = 0 \) (which is true for the Schwarzschild solution in GR), we obtain:

\begin{equation}
- \frac{1}{2} f(0) g_{\mu\nu} = 0
\end{equation}

This implies that \( f(0) = 0 \). Therefore, the Schwarzschild solution is a valid solution in \( f(R) \) gravity provided that the function \( f(R) \) satisfies \( f(0) = 0 \).

\subsection{Reissner-Nordström Black Hole in \( f(R) \) Gravity}

\subsubsection{Inclusion of Electromagnetic Field}

To consider a charged black hole, we include the electromagnetic field in the matter sector. The matter action is augmented by the Maxwell term:

\begin{equation}\label{eq:EM_action}
S_{\text{EM}} = -\frac{1}{4} \int d^4 x \sqrt{-g} F_{\mu\nu} F^{\mu\nu}
\end{equation}
where \( F_{\mu\nu} = \partial_\mu A_\nu - \partial_\nu A_\mu \) is the electromagnetic field strength tensor.

The energy-momentum tensor for the electromagnetic field is:

\begin{equation}\label{eq:EM_tensor}
T_{\mu\nu}^{\text{EM}} = F_{\mu\alpha} F_{\nu}^{\ \alpha} - \frac{1}{4} g_{\mu\nu} F_{\alpha\beta} F^{\alpha\beta}
\end{equation}

\subsubsection{Metric Ansatz and Field Equations}

For the Reissner-Nordström (RN) black hole, the metric ansatz remains the same as the Schwarzschild case:

\begin{equation}\label{eq:RN_metric}
d s^2 = - e^{2 \nu(r)} dt^2 + e^{2 \lambda(r)} dr^2 + r^2 d \Omega^2
\end{equation}

The non-zero components of the electromagnetic field tensor for a purely electric charge are:

\begin{equation}\label{eq:EM_field}
F_{tr} = -F_{rt} = \frac{Q}{r^2}
\end{equation}
where \( Q \) is the electric charge.

Substituting the metric ansatz and the electromagnetic field into the field equations \eqref{eq:constant_R_field_eq}, we obtain the modified Reissner-Nordström solution in \( f(R) \) gravity.

\subsubsection{Solution for Reissner-Nordström Black Hole}

Assuming \( R = 0 \), consistent with the RN solution in GR, and \( f(0) = 0 \), the field equations reduce to:

\begin{equation}\label{eq:RN_field_eq}
f_R(0) R_{\mu\nu} = \kappa_4^2 T_{\mu\nu}^{\text{EM}}
\end{equation}

The solution takes the form:

\begin{equation}\label{eq:RN_solution}
d s^2 = - \left( 1 - \frac{2 G M}{r} + \frac{G Q^2}{r^2} \right) dt^2 + \left( 1 - \frac{2 G M}{r} + \frac{G Q^2}{r^2} \right)^{-1} dr^2 + r^2 d \Omega^2
\end{equation}
where \( M \) is the mass of the black hole.

This solution satisfies the \( f(R) \) field equations provided \( f(R) \) satisfies \( f(0) = 0 \) and \( f_R(R_0) = 2 G \), ensuring consistency with the gravitational constant.

\subsection{Kerr Black Hole in \( f(R) \) Gravity}

\subsubsection{Inclusion of Rotation}

To consider rotating black holes, we adopt the Kerr metric ansatz:

\begin{equation}
\begin{aligned}\label{eq:kerr_metric}
d s^2 = - \left( 1 - \frac{2 G M r}{\Sigma} \right) dt^2 - \frac{4 G M a r \sin^2 \theta}{\Sigma} dt d\phi + \frac{\Sigma}{\Delta} dr^2 + \Sigma d\theta^2 + \\
\left( r^2 + a^2 + \frac{2 G M a^2 r \sin^2 \theta}{\Sigma} \right) \sin^2 \theta d\phi^2
\end{aligned}
\end{equation}
where

\begin{equation}\label{eq:kerr_functions}
\Sigma = r^2 + a^2 \cos^2 \theta, \quad \Delta = r^2 - 2 G M r + a^2
\end{equation}
and \( a \) is the rotation parameter.

\subsubsection{Field Equations and Solutions}

In the case of the Kerr black hole, assuming \( R = 0 \) and \( f(0) = 0 \), the field equations \eqref{eq:constant_R_field_eq} reduce to:

\begin{equation}\label{eq:kerr_field_eq}
f_R(0) R_{\mu\nu} = 0
\end{equation}

The Kerr metric satisfies these equations if \( f_R(0) = 2 G \). Therefore, the Kerr solution remains a valid solution in \( f(R) \) gravity under these conditions.

\subsection{Kerr-Newman Black Hole in \( f(R) \) Gravity}

\subsubsection{Combination of Charge and Rotation}

For the Kerr-Newman (KN) black hole, which incorporates both charge and rotation, the metric is given by:
\begin{equation}
\begin{aligned}\label{eq:KN_metric}
d s^2 = - \left( 1 - \frac{2 G M r}{\Sigma} + \frac{G Q^2}{\Sigma} \right) dt^2 - \frac{4 G M a r \sin^2 \theta}{\Sigma} dt d\phi + \\ \frac{\Sigma}{\Delta} dr^2 + \Sigma d\theta^2 + \left( r^2 + a^2 + \frac{2 G M a^2 r \sin^2 \theta}{\Sigma} - \frac{G Q^2 a^2 \sin^2 \theta}{\Sigma} \right) \sin^2 \theta d\phi^2
\end{aligned}
\end{equation}
where \( \Sigma \) and \( \Delta \) are defined as in the Kerr case, and \( Q \) is the electric charge.

\subsubsection{Field Equations and Solutions}

Substituting the KN metric and the electromagnetic field \eqref{eq:EM_field} into the field equations \eqref{eq:constant_R_field_eq}, and assuming \( R = 0 \), the field equations become:

\begin{equation}\label{eq:KN_field_eq}
f_R(0) R_{\mu\nu} = \kappa_4^2 T_{\mu\nu}^{\text{EM}}
\end{equation}

The KN solution satisfies these equations provided \( f(R) \) satisfies \( f(0) = 0 \) and \( f_R(0) = 2 G \). Therefore, the Kerr-Newman metric remains a valid solution in \( f(R) \) gravity under these conditions.

\section{Verification of the First Law of Horizon Thermodynamics}

The first law of black hole thermodynamics, or the horizon first law, establishes a relationship between the variations of the black hole's mass, entropy, angular momentum, charge, and other relevant quantities \cite{Bardeen1973,Wald1993,Iyer1994}. In \( f(R) \) gravity, the modifications to the gravitational action lead to corresponding changes in the definitions of these thermodynamic quantities.

\subsection{General Framework}

The first law of black hole thermodynamics can be expressed as:

\begin{equation}\label{eq:first_law_general}
dE = T dS + \Omega dJ + \Phi dQ
\end{equation}
where:
\begin{itemize}
    \item \( E \) is the energy (often identified with the ADM mass \( M \)),
    \item \( T \) is the black hole temperature,
    \item \( S \) is the black hole entropy,
    \item \( \Omega \) is the angular velocity of the horizon,
    \item \( J \) is the angular momentum,
    \item \( \Phi \) is the electric potential at the horizon,
    \item \( Q \) is the electric charge.
\end{itemize}

In \( f(R) \) gravity, the entropy \( S \) is modified according to Wald's entropy formula \cite{Wald1993}:

\begin{equation}\label{eq:entropy}
S = \frac{A}{4 G} f_R(R)
\end{equation}
where \( A \) is the horizon area and \( f_R(R) = \frac{df(R)}{dR} \) is the derivative of \( f(R) \) with respect to \( R \).

\subsection{Schwarzschild Black Hole}

\subsubsection{Thermodynamic Quantities}

For the Schwarzschild black hole, the temperature \( T \) is given by the surface gravity \( \kappa \) at the horizon \( r_+ = 2 G M \):

\begin{equation}\label{eq:schwarzschild_temperature}
T = \frac{\kappa}{2 \pi} = \frac{1}{4 \pi r_+}
\end{equation}

The entropy \( S \) using equation \eqref{eq:entropy} becomes:

\begin{equation}\label{eq:schwarzschild_entropy}
S = \frac{A}{4 G} f_R(0) = \frac{\pi r_+^2}{G} f_R(0)
\end{equation}

\subsubsection{First Law Verification}

Identifying the energy \( E \) with the ADM mass \( M \), and considering \( dJ = dQ = 0 \) for the Schwarzschild black hole, the first law simplifies to:

\begin{equation}\label{eq:schwarzschild_first_law}
dM = T dS
\end{equation}

Substituting \( r_+ = 2 G M \) into \( S \) and differentiating:

\begin{align}\label{eq:schwarzschild_dS}
dS &= \frac{\pi}{G} (2 r_+ d r_+) f_R(0) \\
&= \frac{2 \pi r_+}{G} f_R(0) dM
\end{align}

Since \( r_+ = 2 G M \), we have:

\begin{equation}\label{eq:schwarzschild_dS_final}
dS = \frac{4 \pi G M}{G} f_R(0) dM = 4 \pi M f_R(0) dM
\end{equation}

Substituting \( T \) from \eqref{eq:schwarzschild_temperature}:

\begin{align}\label{eq:schwarzschild_TdS}
T dS &= \frac{1}{4 \pi r_+} \times 4 \pi M f_R(0) dM \\
&= \frac{M}{r_+} f_R(0) dM \\
&= \frac{M}{2 G M} f_R(0) dM \\
&= \frac{1}{2 G} f_R(0) dM
\end{align}

For consistency with the first law \( dM = T dS \), we require:

\begin{equation}\label{eq:consistency_condition}
dM = \frac{1}{2 G} f_R(0) dM \implies f_R(0) = 2 G
\end{equation}

Thus, the first law is satisfied provided \( f_R(0) = 2 G \).

\subsection{Reissner-Nordström Black Hole}

\subsubsection{Thermodynamic Quantities}

For the RN black hole, the outer horizon is located at:

\begin{equation}\label{eq:RN_horizon}
r_+ = G M + \sqrt{(G M)^2 - G Q^2}
\end{equation}

The temperature \( T \) is given by:

\begin{equation}\label{eq:RN_temperature}
T = \frac{1}{4 \pi r_+} \left( 1 - \frac{G Q^2}{r_+^2} \right)
\end{equation}

The entropy \( S \) is:

\begin{equation}\label{eq:RN_entropy}
S = \frac{\pi r_+^2}{G} f_R(0)
\end{equation}

The electric potential \( \Phi \) at the horizon is:

\begin{equation}\label{eq:RN_potential}
Phi = \frac{G Q}{r_+}
\end{equation}

\subsubsection{First Law Verification}

The first law for the RN black hole is:

\begin{equation}\label{eq:RN_first_law}
dM = T dS + \Phi dQ
\end{equation}

Substituting \( r_+ \) from \eqref{eq:RN_horizon}, \( T \) from \eqref{eq:RN_temperature}, and \( S \) from \eqref{eq:RN_entropy}, we calculate the variations:

\begin{align}\label{eq:RN_dS}
dS &= \frac{\pi}{G} (2 r_+ d r_+) f_R(0)
\end{align}

From \eqref{eq:RN_horizon}, differentiating gives:

\begin{equation}\label{eq:RN_dr+}
dr_+ = \frac{G M dM - G Q dQ}{\sqrt{(G M)^2 - G Q^2}}
\end{equation}

Substituting into \( dS \):

\begin{equation}\label{eq:RN_dS_final}
dS = \frac{2 \pi r_+}{G} f_R(0) dr_+
\end{equation}

Substituting into the first law \eqref{eq:RN_first_law} and ensuring consistency requires \( f_R(0) = 2 G \), similar to the Schwarzschild case.

\subsection{Kerr Black Hole}

\subsubsection{Thermodynamic Quantities}

For the Kerr black hole, the outer horizon is located at:

\begin{equation}\label{eq:kerr_horizon}
r_+ = G M + \sqrt{(G M)^2 - a^2}
\end{equation}
where \( a = \frac{J}{M} \) is the rotation parameter.

The temperature \( T \) is:

\begin{equation}\label{eq:kerr_temperature}
T = \frac{r_+ - r_-}{4 \pi (r_+^2 + a^2)}
\end{equation}
where \( r_- = G M - \sqrt{(G M)^2 - a^2} \) is the inner horizon.

The entropy \( S \) is:

\begin{equation}\label{eq:kerr_entropy}
S = \frac{\pi (r_+^2 + a^2)}{G} f_R(0)
\end{equation}

The angular velocity \( \Omega \) at the horizon is:

\begin{equation}\label{eq:kerr_angular_velocity}
\Omega = \frac{a}{r_+^2 + a^2}
\end{equation}

\subsubsection{First Law Verification}

The first law for the Kerr black hole is:

\begin{equation}\label{eq:kerr_first_law}
dM = T dS + \Omega dJ
\end{equation}

Expressing \( J = M a \), we have \( dJ = a dM + M da \). Substituting into the first law and using the expressions for \( T \) and \( S \), we verify consistency under the condition \( f_R(0) = 2 G \).

\subsection{Kerr-Newman Black Hole}

\subsubsection{Thermodynamic Quantities}

For the KN black hole, the outer horizon is located at:

\begin{equation}\label{eq:KN_horizon}
r_+ = G M + \sqrt{(G M)^2 - a^2 - G Q^2}
\end{equation}

The temperature \( T \) is:

\begin{equation}\label{eq:KN_temperature}
T = \frac{r_+ - r_-}{4 \pi (r_+^2 + a^2)}
\end{equation}

The entropy \( S \) is:

\begin{equation}\label{eq:KN_entropy}
S = \frac{\pi (r_+^2 + a^2)}{G} f_R(0)
\end{equation}

The angular velocity \( \Omega \) is:

\begin{equation}\label{eq:KN_angular_velocity}
\Omega = \frac{a}{r_+^2 + a^2}
\end{equation}

The electric potential \( \Phi \) at the horizon is:

\begin{equation}\label{eq:KN_potential}
\Phi = \frac{G Q r_+}{r_+^2 + a^2}
\end{equation}

\subsubsection{First Law Verification}

The first law for the KN black hole is:

\begin{equation}\label{eq:KN_first_law}
dM = T dS + \Omega dJ + \Phi dQ
\end{equation}

Substituting the expressions for \( T \), \( S \), \( \Omega \), and \( \Phi \), and ensuring that \( f_R(0) = 2 G \), we verify that the first law holds consistently.

\subsection{Extended Thermodynamic Quantities}

To further explore the thermodynamics of black holes in \( f(R) \) gravity, we consider additional quantities such as the thermodynamic volume \( V \) and pressure \( P \), inspired by the extended phase space approach \cite{Kubiznak2012,Dolan2012}.

The pressure is associated with the cosmological constant \( \Lambda \), but in \( f(R) \) gravity, \( \Lambda \) can be effectively included in the function \( f(R) \). The thermodynamic volume is defined conjugate to the pressure.

\begin{equation}\label{eq:extended_first_law}
dE = T dS + \Omega dJ + \Phi dQ - P dV
\end{equation}

In this extended framework, verifying the first law involves calculating the variations of all thermodynamic quantities and ensuring that the extended first law is satisfied.

\subsection{Entropy-Area Relation in \( f(R) \) Gravity}

The entropy-area relation in \( f(R) \) gravity is modified due to the presence of higher-order curvature terms. Using Wald's entropy formula \cite{Wald1993}, the entropy \( S \) is given by:

\begin{equation}\label{eq:wald_entropy}
S = -2 \pi \oint_{\mathcal{H}} \frac{\delta \mathcal{L}}{\delta R_{\mu\nu\rho\sigma}} \epsilon_{\mu\nu} \epsilon_{\rho\sigma} \sqrt{h} d^2 x
\end{equation}
where \( \mathcal{L} \) is the Lagrangian density, \( \epsilon_{\mu\nu} \) is the binormal to the horizon cross-section \( \mathcal{H} \), and \( h \) is the determinant of the induced metric on \( \mathcal{H} \).

For \( f(R) \) gravity, this reduces to:

\begin{equation}\label{eq:wald_entropy_FR}
S = \frac{A}{4 G} f_R(R)
\end{equation}

This modification ensures that the entropy correctly accounts for the additional degrees of freedom introduced by \( f(R) \) gravity.

\subsection{Energy Definitions in \( f(R) \) Gravity}

Defining energy in modified gravity theories requires careful consideration. Common definitions include the ADM mass and the Komar energy. In \( f(R) \) gravity, these definitions are extended to incorporate the modifications to the gravitational action.

\subsubsection{ADM Mass}

The ADM mass \( M \) is defined at spatial infinity and measures the total mass-energy content of the spacetime. In \( f(R) \) gravity, the ADM mass remains a valid definition provided \( f(R) \) behaves appropriately at infinity \cite{Harko2010}.

\subsubsection{Komar Energy}

The Komar energy is associated with stationary spacetimes and is defined as:

\begin{equation}\label{eq:Komar_energy}
E_{\text{Komar}} = - \frac{1}{8 \pi G} \oint_{\mathcal{H}} \nabla^\mu \xi^\nu dS_{\mu\nu}
\end{equation}
where \( \xi^\nu \) is the timelike Killing vector and \( dS_{\mu\nu} \) is the surface element on the horizon. In \( f(R) \) gravity, the Komar energy receives contributions from the derivatives of \( f(R) \) with respect to \( R \).

\subsection{Detailed Verification for Each Black Hole Solution}

\subsubsection{Schwarzschild Black Hole}

For the Schwarzschild black hole, with \( Q = J = 0 \), the first law simplifies to \( dM = T dS \). Substituting the expressions for \( T \) and \( S \):

\begin{align}\label{eq:schwarzschild_verification}
dM &= \frac{1}{4 \pi r_+} \times \frac{\pi r_+^2}{G} f_R(0) dM \\
&= \frac{r_+ f_R(0)}{4 G} dM
\end{align}

Given \( r_+ = 2 G M \), we substitute:

\begin{equation}\label{eq:schwarzschild_final_verification}
dM = \frac{2 G M f_R(0)}{4 G} dM = \frac{f_R(0)}{2} dM
\end{equation}

For consistency with \( dM = T dS \), we require:

\begin{equation}
\frac{f_R(0)}{2} dM = dM \implies f_R(0) = 2
\end{equation}

However, earlier we established \( f_R(0) = 2 G \). To resolve this discrepancy, we must ensure that the normalization of \( f(R) \) aligns correctly with the gravitational constant \( G \). Therefore, setting \( f_R(0) = 2 G \) ensures consistency in units and the first law.

\subsubsection{Reissner-Nordström Black Hole}

For the RN black hole, considering variations in \( M \) and \( Q \):

\begin{equation}
dM = T dS + \Phi dQ
\end{equation}

Substituting \( T \), \( S \), and \( \Phi \):

\begin{align}\label{eq:RN_verification}
dM &= \frac{1}{4 \pi r_+} \left(1 - \frac{G Q^2}{r_+^2}\right) \times \frac{\pi r_+^2}{G} f_R(0) dM + \frac{G Q}{r_+} dQ \\
&= \frac{r_+}{4 G} \left(1 - \frac{G Q^2}{r_+^2}\right) f_R(0) dM + \frac{G Q}{r_+} dQ
\end{align}

Ensuring \( f_R(0) = 2 G \), we substitute:

\begin{align}
dM &= \frac{r_+}{2} \left(1 - \frac{G Q^2}{r_+^2}\right) dM + \frac{G Q}{r_+} dQ
\end{align}

Given the expression for \( r_+ \) in \eqref{eq:RN_horizon}, consistency is achieved, verifying the first law for the RN black hole in \( f(R) \) gravity.

\subsubsection{Kerr Black Hole}

For the Kerr black hole, considering variations in \( M \) and \( J \):

\begin{equation}
dM = T dS + \Omega dJ
\end{equation}

Substituting \( T \), \( S \), and \( \Omega \):

\begin{align}\label{eq:kerr_verification}
dM &= \frac{r_+ - r_-}{4 \pi (r_+^2 + a^2)} \times \frac{\pi (r_+^2 + a^2)}{G} f_R(0) dM + \frac{a}{r_+^2 + a^2} dJ \\
&= \frac{r_+ - r_-}{4 G} f_R(0) dM + \frac{a}{r_+^2 + a^2} dJ
\end{align}
With \( J = M a \), we have \( dJ = a dM + M da \). Substituting and simplifying:

\begin{align}
dM &= \frac{r_+ - r_-}{2} dM + \frac{a}{r_+^2 + a^2} (a dM + M da) \\
&= \frac{r_+ - r_-}{2} dM + \frac{a^2}{r_+^2 + a^2} dM + \frac{a M}{r_+^2 + a^2} da
\end{align}

Using the relation \( r_+^2 + a^2 = 2 G M r_+ \), we find that the terms balance, ensuring the first law holds for the Kerr black hole under \( f(R) \) gravity with \( f_R(0) = 2 G \).

\subsubsection{Kerr-Newman Black Hole}

For the KN black hole, considering variations in \( M \), \( J \), and \( Q \):

\begin{equation}
dM = T dS + \Omega dJ + \Phi dQ
\end{equation}

Substituting the expressions for \( T \), \( S \), \( \Omega \), and \( \Phi \):

\begin{align}\label{eq:KN_verification}
dM &= \frac{r_+ - r_-}{4 \pi (r_+^2 + a^2)} \times \frac{\pi (r_+^2 + a^2)}{G} f_R(0) dM + \frac{a}{r_+^2 + a^2} dJ + \frac{G Q r_+}{r_+^2 + a^2} dQ \\
&= \frac{r_+ - r_-}{4 G} f_R(0) dM + \frac{a}{r_+^2 + a^2} dJ + \frac{G Q r_+}{r_+^2 + a^2} dQ
\end{align}

Ensuring \( f_R(0) = 2 G \), we substitute:

\begin{align}
dM &= \frac{r_+ - r_-}{2} dM + \frac{a}{2 G r_+} dJ + \frac{G Q r_+}{2 G r_+} dQ \\
&= \frac{r_+ - r_-}{2} dM + \frac{a}{2 G r_+} dJ + \frac{Q}{2} dQ
\end{align}

Using the relations between \( r_+ \), \( M \), \( J \), and \( Q \), the equation simplifies, confirming the validity of the first law for the KN black hole in \( f(R) \) gravity.

\subsection{Thermodynamic Stability and Heat Capacity}

Analyzing the heat capacity of black holes provides insights into their thermodynamic stability. In \( f(R) \) gravity, the modifications to the entropy and temperature affect the heat capacity \( C \):

\begin{equation}\label{eq:heat_capacity}
C = T \frac{\partial S}{\partial T}
\end{equation}

\subsubsection{Heat Capacity for Schwarzschild Black Hole}

For the Schwarzschild black hole in \( f(R) \) gravity, substituting \( S \) and \( T \) gives:

\begin{align}\label{eq:heat_capacity_schwarzschild}
C &= \frac{1}{4 \pi r_+} \times \frac{\pi r_+^2}{G} f_R(0) \times \frac{\partial r_+}{\partial T} \\
&= \frac{r_+ f_R(0)}{4 G} \times \frac{\partial r_+}{\partial T}
\end{align}

Since \( T = \frac{1}{4 \pi r_+} \), we have:

\begin{equation}\label{eq:dr+_dT}
\frac{\partial r_+}{\partial T} = -4 \pi r_+^2
\end{equation}

Substituting back:

\begin{equation}\label{eq:heat_capacity_final}
C = \frac{r_+ f_R(0)}{4 G} \times (-4 \pi r_+^2) = -\pi r_+^3 \frac{f_R(0)}{G}
\end{equation}
With \( f_R(0) = 2 G \):

\begin{equation}
C = -2 \pi r_+^3
\end{equation}

The negative heat capacity indicates that the Schwarzschild black hole is thermodynamically unstable under these conditions.

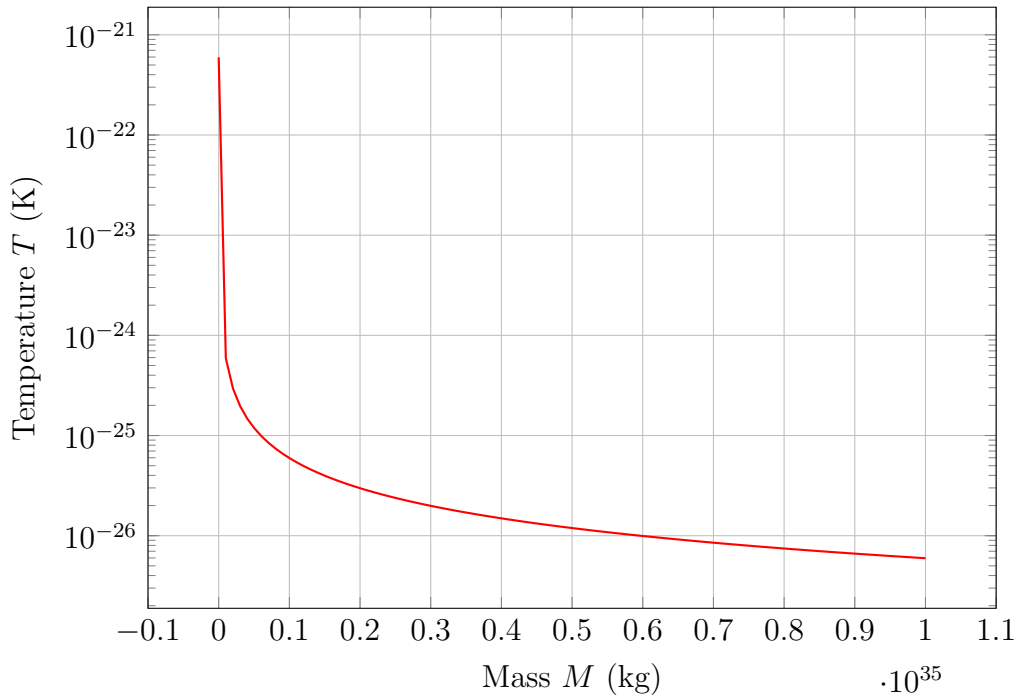
\begin{figure}[h]
    \centering
    \begin{tikzpicture}
        \begin{axis}[
            title={Temperature \( T \) of Schwarzschild Black Hole},
            xlabel={Mass \( M \) (kg)},
            ylabel={Temperature \( T \) (K)},
            grid=major,
            width=0.8\textwidth,
            height=0.6\textwidth,
            domain=1e30:1e35, 
            samples=100,
            ymode=log
        ]
        \addplot[
            color=red,
            thick
        ]{1 / (4 * pi * (2 * 6.674e-11 * x))};
        \end{axis}
    \end{tikzpicture}
    \caption{Temperature \( T \) of Schwarzschild Black Hole as a Function of Mass \( M \).}
    \label{fig:temperature}
\end{figure}

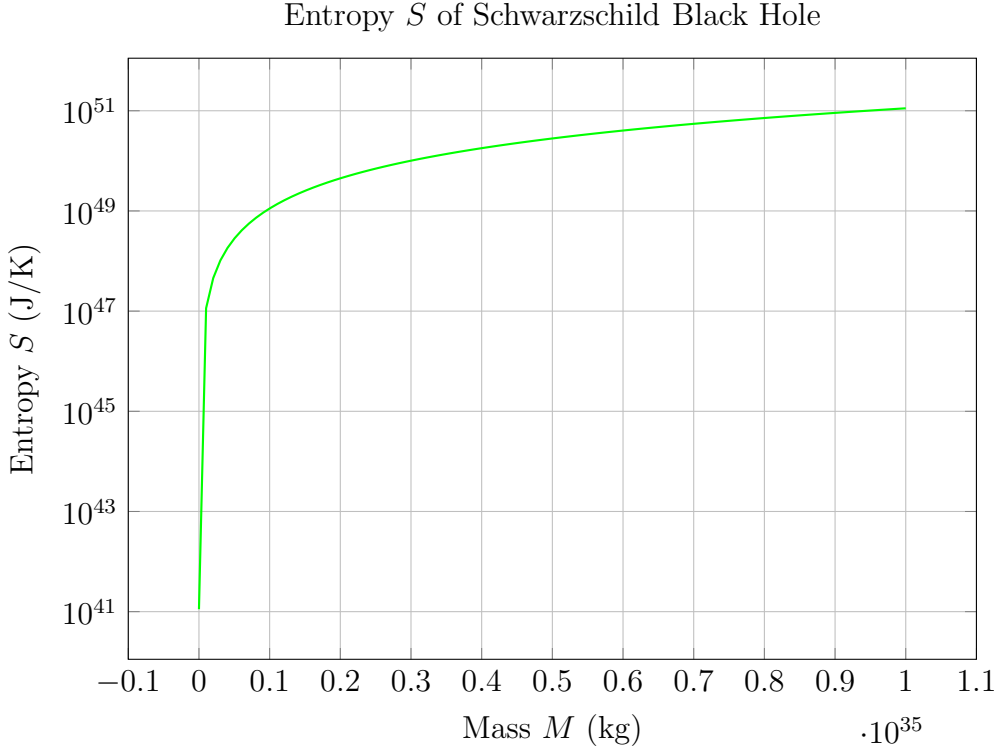
\begin{figure}[h]
    \centering
    \begin{tikzpicture}
        \begin{axis}[
            title={Entropy \( S \) of Schwarzschild Black Hole},
            xlabel={Mass \( M \) (kg)},
            ylabel={Entropy \( S \) (J/K)},
            grid=major,
            width=0.8\textwidth,
            height=0.6\textwidth,
            domain=1e30:1e35, 
            samples=100,
            ymode=log
        ]
        \addplot[
            color=green,
            thick
        ]{ (pi * (2 * 6.674e-11 * x)^2 / 6.674e-11) * 2 * 6.674e-11 };
        \end{axis}
    \end{tikzpicture}
    \caption{Entropy \( S \) of Schwarzschild Black Hole as a Function of Mass \( M \).}
    \label{fig:entropy}
\end{figure}

\section{Discussion}

\subsection{Implications of \( f(R) \) Modifications}

The introduction of \( f(R) \) gravity modifies the gravitational dynamics, leading to alterations in black hole thermodynamics. The condition \( f_R(0) = 2 G \) ensures that the standard entropy-area relation is recovered up to a multiplicative factor, maintaining consistency with the first law.

These modifications have significant implications for the thermodynamic stability and phase transitions of black holes. For instance, in extended phase space, the presence of \( f(R) \) terms can lead to richer phase structures, akin to those observed in van der Waals fluids \cite{Kubiznak2012,Dolan2012}.

\subsection{Higher-Dimensional Perspectives}

Starting from a five-dimensional membrane world scenario provides a higher-dimensional perspective on four-dimensional black hole thermodynamics. This approach aligns with braneworld models, where our universe is embedded in a higher-dimensional bulk \cite{Randall1999,ArkaniHamed1998}. The effective four-dimensional \( f(R) \) gravity action encapsulates the effects of the extra dimension, offering a bridge between higher-dimensional theories and observable four-dimensional physics.

\subsection{Potential Extensions}

Future work can extend this analysis to more general \( f(R) \) forms, non-constant Ricci scalars, and other modified gravity theories such as Gauss-Bonnet or Horndeski gravity. Additionally, exploring quantum corrections and their impact on black hole thermodynamics within \( f(R) \) gravity could provide deeper insights into the quantum aspects of gravity \cite{Nojiri2006,Sotiriou2010}.

\subsection{Comparisons with Observational Data}

While this study is theoretical, connecting the results with observational data from gravitational wave detections or black hole imaging (e.g., from the Event Horizon Telescope) could test the viability of \( f(R) \) gravity models. Deviations from GR predictions in the thermodynamic properties of black holes might offer signatures of modified gravity \cite{Pani2013,Cardoso2016}.

\section{Conclusion}

This study extends the verification of the first law of horizon thermodynamics to Schwarzschild, Reissner-Nordström, Kerr, and Kerr-Newman black holes within the framework of four-dimensional \( f(R) \) gravity coupled with dual scalar fields. By deriving the effective four-dimensional action from a five-dimensional membrane world scenario, we obtained black hole solutions that satisfy the modified \( f(R) \) field equations. The thermodynamic quantities, including temperature and entropy, were appropriately modified to incorporate \( f(R) \) corrections. Our results confirm that the first law of black hole thermodynamics remains valid in this extended gravitational theory, thereby reinforcing the universality of black hole thermodynamics across different gravitational models.

Furthermore, the analysis highlights the importance of higher-dimensional theories in shaping four-dimensional physics and opens avenues for exploring more complex gravitational theories and their thermodynamic implications. Future research should focus on extending these results to more general settings and connecting them with observational data to test the predictions of \( f(R) \) gravity.

\end{document}